\documentclass{article}
\usepackage{waspaa23,amsmath,amssymb,graphicx,url,times}
\usepackage{color}

\usepackage{ifthen}
\usepackage{lipsum}
\usepackage{booktabs}
\usepackage{graphicx}
\usepackage{multirow} 
\usepackage{lineno}
\usepackage{hyperref}
\usepackage{url}
\usepackage{amsfonts}
\usepackage{bm}

\def\vx{{\bm{x}}}
\def\vy{{\bm{y}}}
\def\vz{{\bm{z}}}

\newcommand{\eref}[1]{(\ref{#1})}
\newcommand{\fref}[1]{Fig.~\ref{#1}}
\newcommand{\tref}[1]{Table~\ref{#1}}
\newcommand{\sref}[1]{Sec.~\ref{#1}}

\usepackage[dvipsnames]{xcolor}
\usepackage[skip=6pt]{caption}

\newcommand\myshade{70}
\colorlet{mywholecolor}{MidnightBlue}
\hypersetup{
  linkcolor  = mywholecolor!\myshade!black,
  citecolor  = mywholecolor!\myshade!black,
  urlcolor   = mywholecolor!\myshade!black,
  colorlinks = true,
}

\newcommand\linesubsec[1]{\vspace{0.8mm}\noindent\textbf{#1 --- }}

\title{General Purpose Audio Effect Removal}

\name{Matthew Rice$^*$ \quad Christian J. Steinmetz$^*$ \quad George Fazekas \quad Joshua D. Reiss}
\address{Centre for Digital Music, Queen Mary University of London, UK}

\begin{document}

\ninept
\maketitle

\def\thefootnote{*}\footnotetext{ These authors contributed equally to this work.}\def\thefootnote{\arabic{footnote}}

\begin{sloppy}

\begin{abstract}
Although the design and application of audio effects is well understood, the inverse problem of removing these effects is significantly more challenging and far less studied. 
Recently, deep learning has been applied to audio effect removal; however, existing approaches have focused on narrow formulations considering only one effect or source type at a time. 
In realistic scenarios, multiple effects are applied with varying source content. This motivates a more general task, which we refer to as \emph{general purpose audio effect removal}.
We developed a dataset for this task using five audio effects across four different sources and used it to train and evaluate a set of existing architectures. 
We found that no single model performed optimally on all effect types and sources.
To address this, we introduced RemFX, an approach designed to mirror the compositionality of applied effects. 
We first trained a set of the best-performing effect-specific removal models and then leveraged an audio effect classification model to dynamically construct a graph of our models at inference.
We found our approach to outperform single model baselines, although examples with many effects present remain challenging.

\end{abstract}


\section{Introduction}
\label{sec:intro}

% high-level background
Audio effects are signal processing devices used to shape sonic characteristics and they play a central role in audio production with applications in music, film, broadcast, and video games~\cite{wilmering2020history}. 
While there is a mature body of work for the design and implementation of audio effects~\cite{zolzer2002dafx}, the inverse problem of audio effect removal is more challenging. 
With the rise of music source separation, interest in remixing, manipulating, and re-purposing recorded audio content has continued to grow~\cite{yang2022don, yang2022upmixing}. 
Audio effect removal unlocks further control over remixing content and also facilitates more powerful audio effect style transfer applications~\cite{steinmetz2022style, koo2022music}.
In addition, audio effect removal also has applications for data generation, which could improve source separation and automatic mixing systems~\cite{steinmetz2021deep, martinez2022FxNormAutomix}, and could also be useful in educational contexts, enabling students to better understand the techniques of professional audio engineers.

% limitations of current approaches
Previous systems for audio effect removal rely on traditional signal processing methods that target specific effects such as distortion~\cite{zavivska2020survey, bernardini2019towards}, compression~\cite{gorlow2013model}, and reverberation~\cite{lebart2001new}. However, these approaches require specialized techniques for each effect and make strong assumptions about the effect implementation, limiting their generality. More recently, deep learning has been applied to this task, enabling a more general and powerful data-driven approach. Nonetheless, existing systems are still narrow in their scope, addressing only a small number of effects such as distortion~\cite{imort2022distortion, moliner2022solving} or reverberation~\cite{saito2022unsupervised, murata2023gibbsddrm}. 
Although some work on speech enhancement has considered the removal of audio effects, which can be seen as corruptions of speech~\cite{serra2022universal, su2020hifi}, these approaches are limited in that they consider only speech and operate at relatively low sample rates ($\leq$ 16 kHz). This limits their applicability in post-production where high-fidelity and support for a wide range of content is required.
In addition, previous work has focused on removing only one effect at a time, whereas real-world audio often has multiple effects present simultaneously~\cite{case2012sound}. It is common to chain together multiple audio effects to achieve a specific result, which significantly complicates the task of removing these effects. 

% contributions
We have three main contributions. Firstly, we address the shortcomings of previous research by introducing a more comprehensive task we name \emph{general purpose audio effect removal}. Secondly, we conduct a series of experiments with our benchmark datasets on single and multiple effect removal. We discover that some architectures are more effective at removing certain effects and that certain effects are more challenging than others. 
We also find that when using single models for multiple effect scenarios, performance is degraded. Thirdly, to overcome this, we introduce RemFX, which we demonstrate surpasses baselines by dynamically composing pretrained effect-specific models at inference.
Despite improved performance, our results suggest more work is necessary in cases with many effects applied at the same time. 
We provide listening examples, datasets, code, and pretrained models to aid further research.\footnote{\url{https://csteinmetz1.github.io/RemFX}}

\begin{figure}[t]
    \centering
    \includegraphics[width=\linewidth,trim={1.2cm 0.0cm 0.3cm 0.0cm},clip]{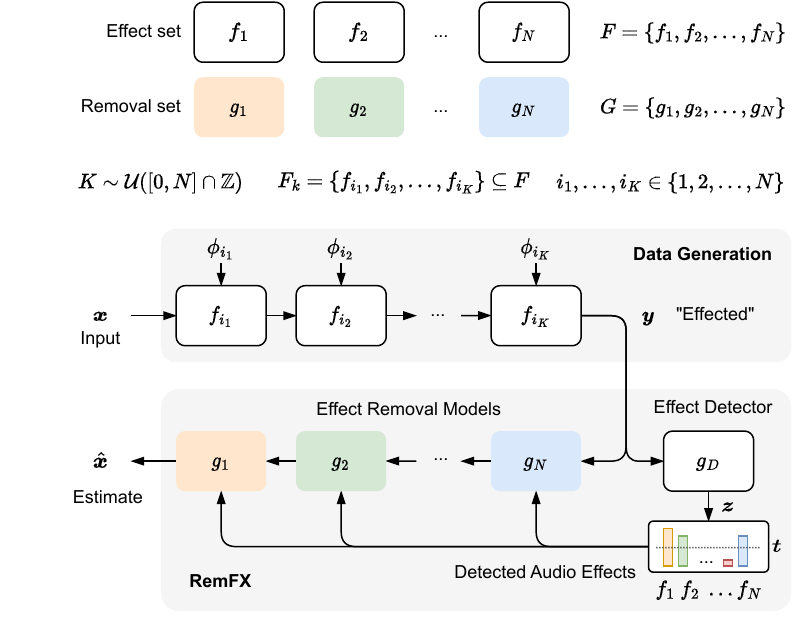}
    %\vspace{0.2cm}
    \caption{ We introduce the task of \emph{general purpose audio effect removal}, which considers removing multiple audio effects from the same recording and we propose RemFX, a compositional approach that dynamically combines effect-specific removal models.}
    \label{fig:remfx}
\end{figure}

\section{Audio effect removal}\label{sec:audio-effect-removal}

Audio effects can be represented by a function
\begin{equation}
\vy = f(\vx ; \phi),
\end{equation}
where $\vx \in \mathbb{R}^T$ denotes the monophonic audio input with $T$ samples, while $\phi \in \mathbb{R}^P$ represents the device operation with $P$ parameters. The function $f$ takes $\vx$ as input and produces an output signal $\vy \in \mathbb{R}^T$, which we refer to as the ``effected'' signal.

In the basic audio effect removal task, the objective is to construct a function $g$ that estimates the signal $\hat{\vx}$ given the effected $\vy$
\begin{equation}\label{eq:simple}
\hat{\vx} = g(\vy).
\end{equation}
While knowledge of the device parameters $\phi$ would be helpful, they are generally unknown and therefore not considered. It should be noted that achieving a perfect reconstruction of the original signal is generally not possible due to the uncertainty surrounding the ground truth given only the effected signal without access to the original device or its parameters. Therefore, our aim is to reduce the perceptual difference between the signals such that when a listener hears the output $\hat{\vx}$, they perceive it as close to the original signal $\vx$.

In audio production, there is a range of effects including time-based effects such as reverberation and delay, dynamic processing such as distortion and compression, spectral effects such as equalization, and modulation effects such as chorus, tremolo, flanger, and phaser~\cite{zolzer2002dafx}. To address this, we represent a set of $N$ audio effects as a set of functions $F = \{ f_1, f_2, ..., f_N \}$ and aim to construct a removal function $g$
\begin{equation}\label{eq:multiple}
\hat{\vx} = g(f_i(\vx; \phi_i)),
\end{equation}
that can recover an estimate of the original signal $\hat{\vx}$ after the application any effect $f_i$ for $i \in {1, 2, ..., N}$. However, this task is further complicated by the fact that multiple effects can be applied to the same recording in order to achieve a desired sound~\cite{case2012sound}. This can be represented by a composition of multiple effect functions, each with its own control parameters. It is important to note that the order of effects can vary and that each effect may or may not be present in any given example, which further complicates the task.

Motivated by this, we formulate the task of \emph{general purpose audio effect removal}. 
We begin by defining a set of $N$ functions $F = \{f_1, f_2, ..., f_N \}$ that represent a group of common audio effects. 
We also define a dataset ${\cal D}=\{\vx^{(j)}\}_{j=1}^{J}$ containing $J$ clean audio recordings where no effects have been applied.  
To generate effected recordings, we randomly sample the number of effects to apply as $K \sim \mathcal{U}([0,N] \cap \mathbb{Z})$. 
Then, we sample $K$ functions without replacement from $F$ to produce a subset $F_K = \{ f_{i_1}, f_{i_2}, \dots, f_{i_K} \} \subseteq F$, where $i_1, i_2, ..., i_K$ represent the indices of elements in $F$.
We then compose the effects in $F_K$ following the order in which they were drawn and sample continuous parameters $\phi_{i_k} \sim \mathcal{U}(a_{i_k},b_{i_k})$ for the $k$-th effect over a predefined range $[a_{i_k} ,b_{i_k}]$. 
We represent the composition of these $K$ functions with random parameters as
\begin{equation}\label{eq:general}
    \vy^{(j)} = f_K(f_{K-1}(\dots f_2(f_1(\vx^{(j)}; \phi_{i_1}); \phi_{i_2}) \dots ; \phi_{i_{K-1}}); \phi_{i_K}),
\end{equation}
where $\vy^{(j)}$ is the resulting output of processing $\vx^{(j)}$, the $j$-th example from the dataset. 
Our goal is to construct a function $g$ such that, given a signal processed by a randomly sampled composition of effects, it will produce an estimate of the recording $\hat{\vx}^{(j)}$ without the presence of effects, minimizing a loss function $\mathcal{L}(\hat{\vx}^{(j)}, \vx^{(j)} )$.
While in some cases effects may be applied in parallel or using more complex routing, our formulation that considers sequential effects captures much of the complexity in real-world audio effect removal.

\newpage
\section{Approach}\label{sec:approach}

As introduced in \sref{sec:audio-effect-removal}, the \emph{general purpose audio effect removal} task involves removing any number of audio effects applied to a recording from a set of possible effects. 
%These effects can be applied in any order and any number of effects from the set can be applied.
One straightforward approach to address this task could involve using a single neural network model to remove all effects at once with the model trained to regress the original signal. 
We refer to these approaches as monolithic networks since they use a singular model to remove a range of effects.
However, due to the combinatorial and compositional nature of the audio effect removal task, we hypothesize that using a monolithic network will not produce adequate results.
Given $N$ different effects and assuming that each effect is applied at most once, the total number of possible effect configurations is given by
$\sum_{k=0}^{N} P(N,k) = \sum_{k=0}^{N} \frac{N!}{(N-k)!}$, 
the sum of all permutations across each number of chosen effects $k 
\in {1, 2, ..., N}$.
Beyond the combinatorial nature of the problem, it is also likely that there will be a significant variance in the difficulty of training examples since some examples will contain $N$ effects while others may contain none. 
This may lead the network to focus more on difficult examples that contribute to higher training error. Furthermore, this may disrupt training, potentially harming performance on simpler cases. 

\renewcommand{\arraystretch}{0.9}
\setlength{\tabcolsep}{5.5pt}
\begin{table*}[t]
    \centering
    \begin{tabular}{l r c c c c c c c c c c c c c} \toprule
        \multirow{2}{*}{Approach} & \multirow{2}{*}{Params} & \multicolumn{2}{c}{DST} & \multicolumn{2}{c}{DRC} & 
                                  \multicolumn{2}{c}{RVB} & \multicolumn{2}{c}{CHS} & \multicolumn{2}{c}{DLY}  & \multicolumn{2}{c}{AVG} \\ \cmidrule(lr){3-4} \cmidrule(lr){5-6} \cmidrule(lr){7-8} \cmidrule(lr){9-10} \cmidrule(lr){11-12} \cmidrule(lr){13-14}
                  &          & SI-SDR   & STFT  & SI-SDR    & STFT  & SI-SDR    & STFT  & SI-SDR    & STFT  & SI-SDR    & STFT  & SI-SDR    & STFT  \\ \midrule
        Input     &   -      & 16.37    & 0.654 & 15.57     & 0.779 & 9.30      & 0.866 & 8.31      & 0.539 & 11.28     & 0.742 & 12.17     & 0.716  \\ \midrule
        DPTNet    &  2.9\,M  & 22.38    & 0.798 & 16.95     & 0.810 & 9.817     & 1.128 & 8.50      & 0.870 & 11.768    & 0.957 & 13.89     & 0.913 \\      
        UMX       &  6.3\,M  & 17.38    & 0.505 & 15.39     & 0.534 & 11.39     & 0.706 & 8.88      & 0.534 & 12.87     & 0.688 & 13.18     & 0.593 \\
        DCUNet    &  7.7\,M  & 16.27    & 0.528 & 13.80     & 0.591 & 12.13     & \textbf{0.645} & \textbf{11.08}     & \textbf{0.504} & 13.48     & \textbf{0.616} & 13.35     & 0.577 \\      
        TCN       &  10.0\,M & 18.47    & 0.632 & 14.49     & 0.733 & 13.25     & 0.804 & 8.452     & 0.669 & 11.23     & 0.882 & 13.18     & 0.744 \\       
        HDemucs   &  83.6\,M & \textbf{24.36}    & \textbf{0.402} & \textbf{20.08}     & \textbf{0.422} & \textbf{13.59}     & 0.735 & 9.828     & 0.580 & \textbf{13.54}     & 0.671 & \textbf{16.30}     & \textbf{0.562} \\ 
        \bottomrule
    \end{tabular}
    \caption{SI-SDR ($\uparrow$) and Multi-resolution STFT error ($\downarrow$) for effect-specific models trained to remove one effect across five architectures.}
    \label{tab:single-effect}
    \vspace{-0.4cm}
\end{table*}

\setlength{\tabcolsep}{3.3pt}
\begin{table}[t]
    \centering
    \begin{tabular}{l l c c c c c c c} \toprule
         Approach       & Train     & DST     & DRC    & RVB    & CHS    & DLY    &   AVG \\\midrule
\multirow{2}{*}{HDemucs}& Single    & \textbf{0.232} & \textbf{0.377} & 0.124 & -0.061 & 0.069 & \textbf{0.149}\\
                        & Multiple  & 0.099 & 0.217 & 0.031 & -0.121 & -0.048 & 0.036 \\ \midrule
\multirow{2}{*}{DCUNet} & Single    & 0.096 & 0.202 & \textbf{0.212} & \textbf{0.027} & \textbf{0.127} & 0.133 \\
                        & Multiple  & 0.024 & 0.109 & 0.065 & -0.067 & 0.026 & 0.032\\
        \bottomrule
    \end{tabular}
    \caption{Multi-resolution STFTi ($\uparrow$) for single effect models compared to the same architecture trained to remove all five effects.}
    \label{tab:all-effects-one-at-a-time}
    \vspace{-0.3cm}
\end{table}

\subsection{Compositional audio effect removal}
To address the limitations of monolithic networks in this task, we propose a compositional approach, which we name RemFX.
As shown in \fref{fig:remfx}, our approach is designed to mirror the process of applying a series of audio effects.
We achieve this by first constructing a set of $N$ audio effect-specific removal models $G = \{g_1, g_2, \dots, g_N \}$.
We choose the best-performing model architecture for each effect removal model based on our initial experiments. 
We train each of these networks with a separate dataset to remove a different effect from the effect set $F = \{f_1, f_2, \dots, f_N \}$.

After constructing our set of removal models $G$, we then introduce an audio effect detection network $\vz = g_D(\vy)$ where $\vz 
\in \mathbb{R}^{N}$.
This network is trained in a separate task to detect the presence of any effect from $F$ in the effected recording $\vy$, which we frame as a multi-label classification task. 
At inference, we apply a threshold $t$ to the logits $\vz = (z_1, z_2, ..., z_N)$ from $g_D$, selecting all effects where $\vz \geq t$. 
This enables us to construct a series connection of our effect-specific removal models from $G$ and then apply this composite function to remove any effects, dynamically adapting computation at inference. We do not estimate the order since we found that random ordering performs similarly to the true ordering (\sref{sec:general}).

%Since our approach requires removing effects one-at-a-time from signals with multiple effects present, any one effect-specific model will encounter audio containing other audio effects. 
During inference, our effect-specific models will encounter effects in the input signal that they are not trained to remove. To improve robustness in these scenarios, we propose an approach called FXAug. 
When training an effect-specific removal model $g_n$ to remove an effect $f_n$ from the effect set $F$, we apply additional distractor effects from the set $F \setminus \{ f_n \}$.
In general, we sample $K_d \sim \mathcal{U}([0,N-1] \cap \mathbb{Z})$ distractor effects and apply them with randomly sampled parameters before applying the effect to be removed $f_n$. We then use the intermediate signal containing the distractor effects as the target signal during training, instead of the clean signal.

Compared to monolithic approaches, our approach offers several benefits: it allows for adaptive computation during inference, running only the removal networks of effects that are present, enables expansion to more effects without requiring complete retraining of existing removal models, and facilitates using different architectures specialized for the removal of each effect type. 

\newpage
\section{Experimental Setup}
\label{sec:exp}

\linesubsec{Dataset}
%One of the challenges in audio effect removal is sourcing sufficient data that has not been processed with audio effects. 
We source audio from four datasets: VocalSet~\cite{wilkins2018vocalset} for singing voice, GuitarSet~\cite{xi2018guitarset} for acoustic guitar, DSD100~\cite{dsd100} for bass guitar, and IDMT-SMT-Drums~\cite{dittmar2014drums} for drums. 
We split each set into train, validation, and test, ensuring there is no overlap between song, performer, or instruments, where applicable. 
We resample to $f_s = 48$\,kHz and split audio into $\sim$5.5\,sec chunks (262144 samples). 
We fix the number of train, validation, and test examples to 8\,k, 1\,k, and 1\,k for each experimental configuration. 
We generate effected audio by applying randomly sampled effects and parameters using \texttt{Pedalboard}~\cite{sobot2023pedalboard}, following \sref{sec:audio-effect-removal}. Parameter ranges are selected heuristically to model real-world use cases. After each effect, we loudness normalize the audio with a target of $-20$\,dB LUFS~\cite{steinmetz2021pyloudnorm}.
We consider five effects: Distortion (DST), Dynamic range compression (DRC), Reverberation (RVB), Chorus (CHS), and Feedback delay (DLY).
This results in 12.1\,h for training, 1.5\,h for validation, and 1.5\,h for testing per experiment.

\linesubsec{Removal models}
We consider five audio processing model architectures in our experiments. 
These include, Hybrid Demucs~\cite{defossez2021hybriddemucs}, DCUNet~\cite{choi2019dcunet}, DPTNet~\cite{chen2020dptnet}, TCN~\cite{rethage2018wavenet, steinmetz2021tcn}, and UMX~\cite{stoter2019open}.

\linesubsec{Detection models}
Similar to past work in effect classification~\cite{stein2010automatic, jurgens2020recognizing, comunita2020guitar}, we consider convolutional architectures operating on Mel spectrograms. 
As baselines, we train single linear layers on top of a set of frozen pretrained audio representations, including PANNs~\cite{kong2020panns}, wav2vec2.0~\cite{baevski2020wav2vec}, and Wav2CLIP~\cite{wu2022wav2clip} ($*$ in Table \ref{tab:detection}). For comparison, we also train PANNs from scratch at $f_s = 48$\,kHz.

\linesubsec{Training}
All models are trained with the Adam optimizer. Removal models are trained for 50\,k steps with an initial learning rate of $10^{-4}$ and weight decay of $10^{-3}$ using a batch size optimal for each model on a single A100 GPU.
Audio effect classifiers are trained with a learning rate of ${ 3 \cdot 10^{-4}}$ for 300 epochs using a batch size of 64.
In both cases, we use learning rate scheduling during training, decreasing by a factor of 10 at 80\% and 95\% through training, and gradient clipping with a value of 10.
While classifiers are trained with binary cross-entropy, removal models are trained with a sum of two terms ${\cal L} = \alpha {\cal L}_{\text{L1}} + \beta {\cal L}_{\text{MR-STFT}}$, with $\alpha = 100$ and $\beta = 1$, where ${\cal L}_{\text{L1}}$ is the L1 distance in the time domain and ${\cal L}_{\text{MR-STFT}}$ is the multi-resolution magnitude STFT loss~\cite{ yamamoto2020parallel,steinmetz2020auraloss}.

\setlength{\tabcolsep}{4.0pt}
\begin{table}[t]
    \centering
    \begin{tabular}{l c c c c c c} \toprule
        Approach & DST & DRC & RVB & CHS & DLY & AVG \\ \midrule
        %VGGish*~\cite{hershey2017cnn}        & \\
        wav2vec2*~\cite{baevski2020wav2vec}  & 0.720 & 0.710 & 0.776 & 0.651 & 0.662 & 0.704 \\
        Wav2CLIP*~\cite{wu2022wav2clip}      & 0.642 & 0.667 & 0.850 & 0.697 & 0.699 & 0.720 \\ 
        PANNs*~\cite{kong2020panns}          & 0.681 & 0.681 & 0.841 & 0.705 & 0.730 & 0.732 \\ \midrule
        PANNs                                & \textbf{0.780} & 0.771 & 0.791 & 0.724 & 0.680 & 0.750 \\
        \phantom{0} + SpecAug                & \textbf{0.780} & \textbf{0.807} & \textbf{0.845} & \textbf{0.751} & \textbf{0.743} & \textbf{0.786} \\
        \bottomrule
    \end{tabular}
    \caption{Class-wise accuracy for the audio effect detection task. %\\ Models with * denote frozen pretrained models where liner layers were trained on representations for the audio effect detection task.
    }
    \label{tab:detection}
    \vspace{-0.42cm}
\end{table}

%\iffalse
\renewcommand{\arraystretch}{0.9}
\setlength{\tabcolsep}{4.2pt}
\begin{table*}[!t]
    \centering
            \vspace{-0.2cm}
    \begin{tabular}{l r c  c c c c c c c c c c c c} \toprule
        \multirow{2}{*}{Approach} &  \multirow{2}{*}{Params.} &   \multicolumn{2}{c}{$N=0$}  & \multicolumn{2}{c}{$N=1$}    & \multicolumn{2}{c}{$N=2$}     & \multicolumn{2}{c}{$N=3$}       & \multicolumn{2}{c}{$N=4$} & \multicolumn{2}{c}{$N=5$} \\ \cmidrule(lr){3-4} \cmidrule(lr){5-6} \cmidrule(lr){7-8} \cmidrule(lr){9-10} \cmidrule(lr){11-12} \cmidrule(lr){13-14} 
                          &                     & SI-SDR   & STFT   & SI-SDR   & STFT              & SI-SDR & STFT                & SI-SDR  & STFT                  & SI-SDR   & STFT           & SI-SDR & STFT             \\ \midrule
        Input             &     -               & Inf      & 0.000  & 11.52    & 0.689             & 6.24   & 1.131               & 3.29    & 1.508                 & 1.31     & 1.799          & -0.33     & 2.058 \\ \midrule
        DCUNet                  &  7.7\,M       & 18.53    & 0.467  & 11.16    & 0.743             & 7.87   & 0.945               & 5.42    & 1.121                 & 3.64     & 1.265          & 2.10      & 1.462 \\ 
        HDemucs-M               & 84\,M         & 19.72    & 0.415  & 11.28    & 0.728             & 8.01   & 0.931               & 5.77    & 1.100                 & 4.29     & 1.223          & 2.10      & 1.337 \\ 
        HDemucs-L               & 334\,M        & 20.78    & 0.410  & 11.53    & 0.725             & 8.17   & 0.924               & 6.08    & 1.084                 & 4.63     & 1.212          & 3.38      & 1.328 \\ 
        HDemucs-XL              & 751\,M        & 20.31    & 0.406  & 11.54    & 0.713             & 8.32   & 0.902               & 6.19    & 1.064                 & 4.73     & 1.190          & 3.38      & \textbf{1.312} \\ \midrule            
        RemFX Oracle         & $\leq$192\,M       & \textbf{Inf}      & \textbf{0.000}  & \textbf{16.99}    & \textbf{0.486}             & \textbf{10.91}  & \textbf{0.762}               & \textbf{7.51}    & \textbf{0.994}                 & \textbf{5.40}     & \textbf{1.170}          & \textbf{3.47}      & 1.360 \\            
        RemFX All             & 192\,M        & 21.99    & 0.234  & 10.26    & 0.841             &  8.44  & 0.939               & 6.46    & 1.084                 & 4.71     & 1.225          & 2.99      & 1.418 \\ %\cmidrule(lr){2-14}
        RemFX Detect          & $\leq$192\,M       & 87.54    & 0.068  & 16.67    & 0.495             & 10.47   & 0.786               & 6.96    & 1.050                 & 4.80     & 1.247          & 2.61      & 1.486 \\
        \bottomrule
         %& *192\,M   & Oracle & Random   & 16.99    & 0.486             & 10.48  & 0.783               & 7.08    & 1.030                 & 4.87     & 1.220          & 2.99      & 1.418 \\
        %& *192\,M   & Oracle & Fixed    & 16.99    & 0.486             & 10.69  & 0.773               & 7.26    & 1.016                 & 5.14     & 1.198          & 3.14      & 1.396 \\ \cmidrule(lr){2-14}
        %& 192\,M    & All    & Random   & 11.52    & 0.689             &  8.44  & 0.939               & 6.46    & 1.084                 & 4.71     & 1.225          & 2.99      & 1.418 \\
        %& *192\,M   & Detect & Random   &          &  \\
    \end{tabular}
    \caption{SI-SDR ($\uparrow$) and MR-STFT error ($\downarrow$) in \emph{general purpose audio effect removal} across fixed number of audio effects $N$. }
    \label{tab:all-effects-multiple-at-a-time}
        \vspace{-0.4cm}
\end{table*}
%\fi

\section{Experiments \& Results}
\vspace{-0.1cm}
\subsection{Effect-specific models}
We report SI-SDR~\cite{le2019sdr} for performance in the time domain and the multi-resolution STFT error for performance in the magnitude frequency domain~\cite{ yamamoto2020parallel,steinmetz2020auraloss}. 
We use ``STFT'' as shorthand to denote the multi-resolution STFT error. 
In some cases, we also report SI-SDRi and STFTi, which indicates the improvement in each metric in comparison to the input signal.
We train one model for each architecture across five effects, resulting in a total of 25 models for the task in \eqref{eq:simple}. 
As shown in \tref{tab:single-effect}, we found that no architecture performs optimally across all removal tasks. 
Hybrid Demucs outperforms others in distortion and compression, whereas DCUNet performs better on chorus. 
Although the performance is similar for reverberation and delay, STFT error suggests that DCUNet performs better while SI-SDR scores are close. 
This aligns with our informal listening, however, it also reveals chorus and delay remain challenging to remove even for the best-performing models.
\vspace{-0.30cm}
\subsection{Monolithic removal models}
As a first step towards the \emph{general purpose audio effect removal} task, we train Hybrid Demucs and DCUNet as monolithic models to remove multiple effects when only one effect is present at a time, as in \eref{eq:multiple}.  
We report the results in \tref{tab:all-effects-one-at-a-time}, comparing the performance to the effect-specific models from the previous experiment.
When training to remove multiple types of effects, we observe that both architectures perform worse as compared to when they are trained to remove only a single effect. This provides evidence for our claim in \sref{sec:approach} that monolithic models may not produce adequate results.

\vspace{-0.5cm}
\subsection{Audio effect detection}
We frame the audio effect detection task as $N$ separate binary classification tasks, where one linear layer followed by a sigmoid generates a prediction for the presence of each effect. 
We report the class-wise accuracy on held-out data in \tref{tab:detection}.
We found superior performance training PANNs from scratch as compared to adapting the pretrained models, and we observed a small benefit (+3.6\% accuracy) from SpecAugment.
However, our results indicate the multiple effect detection task could be further improved, as the best-performing model achieves 78.6\% accuracy across all effects. 

\setlength{\tabcolsep}{6.0pt}
\begin{table}[t]
    \centering
    \begin{tabular}{l l c c c c} \toprule
        \multirow{2}{*}{Effect} & \multirow{2}{*}{Approach} & \multicolumn{2}{c}{Single Effect} &  \multicolumn{2}{c}{w/ Distractors} \\ \cmidrule(lr){3-4} \cmidrule(lr){5-6} 
                                &                   & SI-SDR & STFT & SI-SDR & STFT \\  \midrule
        \multirow{3}{*}{AVG}    & Input             & 12.17 & 0.716 & 10.60 & 0.692    \\ \cmidrule{2-6}
                                & Single Effect     & 16.28 & 0.612 & 14.71 & 0.580    \\ 
                                & + FXAug           & \textbf{16.59} & \textbf{0.554} & \textbf{16.61} & \textbf{0.514}    \\  
        \bottomrule
        
    \end{tabular}
    \caption{Average SI-SDR ($\uparrow$) and MR-STFT error ($\downarrow$) across all effects for single effect removal trained with and without FXAug.}
    \label{tab:effect-augmentation}
        \vspace{-0.38cm}
\end{table}

\vspace{-0.2cm}
\subsection{Audio effect augmentation}
We hypothesized that training effect-specific models with only one effect during training would lead to degraded performance. 
To investigate this and the efficacy of our FXAug approach, we trained a Hybrid Demucs model for each effect, with and without FXAug. We evaluated with two test sets: one with only one effect and one with up to four random distractor effects. We report the mean performance across all five effect-specific models in \tref{tab:effect-augmentation}.
First, we confirm that distractor effects harm performance for models trained with only one effect (no FXAug). 
Second, this is remedied by \mbox{FXAug}, which improves performance in the case of distractors, but also in the case of single effects. Therefore, we use effect-specific models trained with FXAug in our final RemFX system. 

\vspace{-0.2cm}
\subsection{General purpose audio effect removal}\label{sec:general}
In our final experiment, we investigated the performance of systems on the \emph{general purpose audio effect removal} task using our set of audio effects, applying up to five at a time.
We trained monolithic Hybrid Demucs and DCUNet models and compared them against variants of RemFX with results in \tref{tab:all-effects-multiple-at-a-time}. These include All: apply all effect-specific models, Oracle: apply respective models given ground truth labels of the effects that are present, and Detect: use the audio effect classifier to determine these labels.   
For the effect-specific models, we used Hybrid Demucs for distortion and compression, and DCUNet for reverberation, delay, and chorus, along with the best-performing classifier from 
\tref{tab:detection}, using a threshold of $t=0.5$. The ordering of the models was randomized for each example, except for Oracle, which used the ground truth ordering.

\linesubsec{Number of effects}
The case of no effects, $N=0$, exhibits one of the benefits of RemFX, which will not process the input unless audio effects are detected. On the other hand, the monolithic models produced noticeable degradation across both metrics.
For $N=1$, we found that monolithic models struggle, performing worse even than the input across both metrics, while RemFX Oracle achieved a significant improvement. Even RemFX Detect only had a small performance dip ($<2\%$) and still outperformed the monolithic models.
This trend is similar for $N=2$.
While the monolithic models provided a small improvement, RemFX models were superior. 
As the number of effects increases, the gap between RemFX and the monolithic models decreases, as shown in \fref{fig:best-models}. 
While RemFX outperformed the baselines when fewer effects are present, all approaches exhibited degraded performance for $N=4$ and $N=5$, indicating the difficulty of this task. 

\linesubsec{Model scale}
To improve performance of monolithic models, we attempted to further scale Hybrid Demucs. However, we observed minimal improvement in SI-SDR ($\leq$ 1.3\,dB) and STFT ($\leq$ 0.04) across $N$, even when scaling to 751\,M parameters. In comparison, RemFX models performed better for $N \leq 3$ and use fewer parameters, ranging from 0 to 192\,M, depending on the detected effects. 

%XL     20.31   11.54   8.32    6.19    4.73    3.38
% M     19.72   11.28   8.01    5.77    4.29    2.10

\linesubsec{Ordering \& Detection}
For RemFX Oracle, we compared using ground truth and random ordering. We found a slight decrease in SI-SDR ($\leq$ 0.6\,dB) and STFT ($\leq 0.06$) across $N$ when using a random ordering, leading us to conclude the use of random ordering in our RemFX Detect model is an acceptable approach. 
We also established the importance of the classifier, since RemFX All results in a max decrease of 6.41\,dB SI-SDR and 0.346 STFT. 

\begin{figure}[t]
    \centering
    \includegraphics[width=0.95\linewidth, trim={0.2cm 0.3cm 0.3cm 0.2cm},clip]{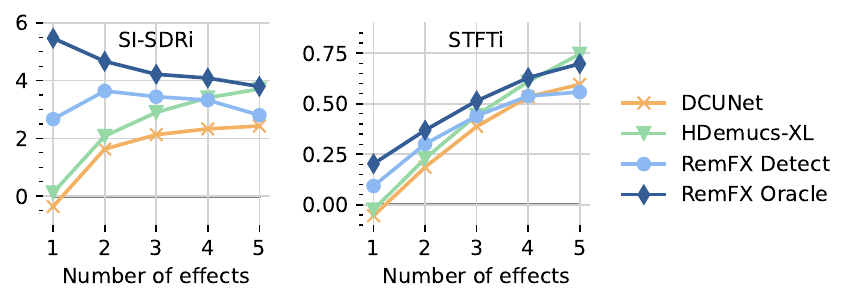}
    \caption{Average SI-SDRi and STFTi across effects.}
    \label{fig:best-models}
    \vspace{-0.4cm}
\end{figure}

\vspace{-0.21cm}
\section{Conclusion}
\vspace{-0.21cm}
\label{sec:conclusion}
We introduced a new task, \emph{general purpose audio effect removal}, and investigated several approaches to tackle it. Our findings suggested that monolithic networks fail to generalize across a varying number of effects; however, our RemFX system yielded improved performance by combining an effect detection model with dynamic construction of effect-specific removal models. 
While the results are promising, our evaluation is limited in that we considered only five effects, each with a single implementation, and without more complex signal routing, such as parallel connections.
Despite these limitations, our method offers promising results in effect removal and provides a direction for improved effect removal systems that are scalable and applicable in real-world scenarios.
We provide code, datasets, and pretrained models to facilitate future work.

% acknowledgement is not included in anon version
\vspace{-0.21cm}
\section{ACKNOWLEDGMENT}
\vspace{-0.21cm}
\label{sec:ack}
CS is supported by the EPSRC UKRI CDT in AI and Music (Grant no. EP/S022694/1). Compute resources provided by Stability AI.

% -------------------------------------------------------------------------
% Either list references using the bibliography style file IEEEtran.bst
\clearpage
\bibliographystyle{IEEEtran}
\bibliography{references}

\end{sloppy}
\end{document}